\documentclass[twoside,twocolumn,9pt]{article}
\usepackage{extsizes}
\usepackage[super,sort&compress,comma]{natbib}
\usepackage[version=3]{mhchem}
\usepackage[left=1.5cm, right=1.5cm, top=1.785cm, bottom=2.0cm]{geometry}
\usepackage{balance}
\usepackage{times,mathptmx}
\usepackage{sectsty}
\usepackage{graphicx}
\usepackage{lastpage}
\usepackage[format=plain,justification=justified,singlelinecheck=false,font={stretch=1.125,small,sf},labelfont=bf,labelsep=space]{caption}
\usepackage{float}
\usepackage{fancyhdr}
\usepackage{fnpos}
\usepackage[english]{babel}
\usepackage{amssymb}
\usepackage{multicol}
\addto{\captionsenglish}{%
  
}
\usepackage{array}
\usepackage{droidsans}
\usepackage{charter}
\usepackage[T1]{fontenc}
\usepackage[usenames,dvipsnames]{xcolor}
\usepackage{setspace}
\usepackage[compact]{titlesec}
\usepackage{hyperref}

\usepackage{epstopdf}

\definecolor{cream}{RGB}{222,217,201}

\definecolor{myc} {rgb} {0,0,0}  

\begin{document}

\pagestyle{fancy}
\thispagestyle{plain}
\fancypagestyle{plain}{

}

\makeFNbottom
\makeatletter
\renewcommand\LARGE{\@setfontsize\LARGE{15pt}{17}}
\renewcommand\Large{\@setfontsize\Large{12pt}{14}}
\renewcommand\large{\@setfontsize\large{10pt}{12}}
\renewcommand\footnotesize{\@setfontsize\footnotesize{7pt}{10}}
\makeatother

\renewcommand{\thefootnote}{\fnsymbol{footnote}}
\renewcommand\footnoterule{\vspace*{1pt}%
\color{cream}\hrule width 3.5in height 0.4pt \color{black}\vspace*{5pt}}
\setcounter{secnumdepth}{5}

\makeatletter
\renewcommand\@biblabel[1]{#1}
\renewcommand\@makefntext[1]%
{\noindent\makebox[0pt][r]{\@thefnmark\,}#1}
\makeatother
\renewcommand{\figurename}{\small{Fig.}~}
\sectionfont{\sffamily\Large}
\subsectionfont{\normalsize}
\subsubsectionfont{\bf}
\setstretch{1.125} 
\setlength{\skip\footins}{0.8cm}
\setlength{\footnotesep}{0.25cm}
\setlength{\jot}{10pt}
\titlespacing*{\section}{0pt}{4pt}{4pt}
\titlespacing*{\subsection}{0pt}{15pt}{1pt}

\fancyfoot{}
\fancyfoot[LO,RE]{\vspace{-7.1pt}}
\fancyfoot[CO]{\vspace{-7.1pt}\hspace{13.2cm}}
\fancyfoot[CE]{\vspace{-7.2pt}\hspace{-14.2cm}}
\fancyfoot[RO]{\footnotesize{\sffamily{1--\pageref{LastPage} ~\textbar  \hspace{2pt}\thepage}}}
\fancyfoot[LE]{\footnotesize{\sffamily{\thepage~\textbar\hspace{3.45cm} 1--\pageref{LastPage}}}}
\fancyhead{}
\renewcommand{\headrulewidth}{0pt}
\renewcommand{\footrulewidth}{0pt}
\setlength{\arrayrulewidth}{1pt}
\setlength{\columnsep}{6.5mm}
\setlength\bibsep{1pt}

\makeatletter
\newlength{\figrulesep}
\setlength{\figrulesep}{0.5\textfloatsep}

\newcommand{\topfigrule}{\vspace*{-1pt}%
\noindent{\color{cream}\rule[-\figrulesep]{\columnwidth}{1.5pt}} }

\newcommand{\botfigrule}{\vspace*{-2pt}%
\noindent{\color{cream}\rule[\figrulesep]{\columnwidth}{1.5pt}} }

\newcommand{\dblfigrule}{\vspace*{-1pt}%
\noindent{\color{cream}\rule[-\figrulesep]{\textwidth}{1.5pt}} }

\makeatother

\twocolumn[
  \begin{@twocolumnfalse}
\vspace{3cm}
\sffamily
\begin{tabular}{m{4.5cm} p{13.5cm} }

& \noindent\LARGE{\textbf{Instabilities in freely expanding sheets of associating viscoelastic fluids}} \\
\vspace{0.3cm} & \vspace{0.3cm} \\

 & \noindent\large{Srishti Arora,\textit{$^{a}$}$\ddag$ Ameur Louhichi,\textit{$^{a,b}$}$\ddag$ Dimitris Vlassopoulos,\textit{$^{b}$} Christian Ligoure,$^{\ast}$\textit{$^{a}$} and Laurence Ramos$^{\ast}$\textit{$^{a}$}} \\

 & \noindent\normalsize{We use the impact of drops on a small solid target as a tool to investigate the behavior of viscoelastic fluids under extreme deformation rates. We study two classes of transient networks: semidilute solutions of supramolecular polymers and suspensions of spherical oil droplets reversibly linked by polymers. The two types of samples display very similar linear viscoelastic properties, which can be described with a Maxwell fluid model, but contrasting nonlinear properties due to different network structure. Upon impact, weakly viscoelastic samples exhibit a behavior qualitatively similar to that of Newtonian fluids: A smooth and regular sheet forms, expands, and then retracts. By contrast, for highly viscoelastic fluids, the thickness of the sheet is found to be very irregular, leading to instabilities and eventually formation of holes. We find that material rheological properties rule the onset of instabilities. We first provide a simple image analysis of the expanding sheets to determine the onset of instabilities. We then demonstrate that a \textcolor{myc} {Deborah number related to the shortest relaxation time associated to the sample structure following a high shear} is the relevant parameter that controls the heterogeneities in the thickness of the sheet, eventually leading to the formation of holes. When the sheet tears-up, data suggest by contrast that the opening dynamics \textcolor{myc} {depends also} on the expansion rate of the sheet. }

\end{tabular}

 \end{@twocolumnfalse} \vspace{0.6cm}

  ]

\renewcommand*\rmdefault{bch}\normalfont\upshape
\rmfamily
\section*{}
\vspace{-1cm}


\footnotetext{\textit{$^{a}$ Laboratoire Charles Coulomb (L2C), Univ. Montpellier, CNRS, Montpellier, France. E-mail: laurence.ramos@umontpellier.fr, christian.ligoure@umontpellier.fr}}
\footnotetext{\textit{$^{b}$ Institute of Electronic Structure and Laser, FORTH, Heraklion 70013, Crete, Greece and Department of Materials Science and Technology, University of Crete, Heraklion 70013, Crete, Greece}}


\footnotetext{\ddag~contribute equally}



\section{Introduction}

It is known that viscoelastic liquids behave as solid-like materials at short times and as viscous liquids at long times. For the simplest case of a material with a unique relaxation time (as the archetype Maxwell fluid), this cross-over time is the sample relaxation time, which can be easily measured in standard linear shear rheology, as long as it falls in the experimentally accessible time window. Depending on the characteristic time scale of the experiments as compared to the intrinsic sample relaxation time, one therefore probes the solid nature or the liquid nature of a sample. This simple physical picture can be exemplified in several experimental contexts, where a viscoelastic fluid is submitted to large and/or fast deformations.
One textbook situation concerns the rupture of fluid materials.
Fracture processes are features associated to the solid-like nature of the materials. Hence if one induces a deformation of the sample at a rate faster than the inverse of its characteristic relaxation time, a viscoelastic material could in principle break as would do a regular solid. Note, however, that alternative mechanisms, such as thermally activated crack nucleation, could also be relevant and anticipate the fracture process.
Fracture of viscoelastic fluids has been observed in various experimental conditions~\cite{ligoure_fractures_2013}, ranging from
shearing a bulk viscoelastic fluid~\cite{berret_evidence_2001,tabuteau_microscopic_2009,skrzeszewska_fracture_2010,ramos_structural_2011}
to stretching a viscoelastic filament and concomitantly visualize the solid-like rupture of a filament with a finite diameter. Such process could occur spontaneously following the free fall of a viscoelastic drop~\cite{smolka_drop_2003,tabuteau_microscopic_2009,tabuteau_ductility_2008}, but more controlled experiments could also be conducted using a filament-stretching rheometer, allowing filament to be stretched at a prescribed elongational rate~\cite{mckinley_filament-stretching_2002}. In this case, when stretched at a sufficiently large rate, transient polymer networks~\cite{tripathi_rheology_2006,arora_brittle_2017} and polymer melts~\cite{huang_multiple_2016,huang_polymer_2017} have been observed to rupture in a solid-like manner, whereas they undergo a liquid-like thinning at smaller rates. In addition, Hele-Shaw experiments allow also a tuning of the deformation imposed to the sample. In these experiments, a fluid pushes a more viscous fluid confined between two plates, leading to viscous fingering. If a viscoelastic fluid is used instead of the viscous fluid, a transition from fingering to fracturing occurs at sufficiently high injection rate, as observed in a wide variety of materials ranging from foams~\cite{hilgenfeldt_foam:_2008,ben_salem_response_2013}, to colloidal suspensions~\cite{Lemaire1991}, and water-based transient networks made of associative polymers~\cite{Zhao1993,IgnesMullol1995} or oil droplets~\cite{mora_saffman-taylor_2010,foyart_fingering_2013}.

Studies to rationalize criteria for the emergence of instabilities in the flow of viscoelastic fluids are often based on a comparison between a sample intrinsic relaxation time and a typical deformation rate for the flow. In this framework, the rupture of polymer liquids in elongational flow has been modeled using a Weissenberg number based on a linear reptation time~\cite{joshi_rupture_2003}. The fracture of dense colloidal suspensions has also been analyzed in terms of critical strain rates compared to structural relaxation times~\cite{smith_fracture_2015}. In the case of Hele-Shaw experiments, on the other hand, detailed quantitative works have been conducted to define the critical injection rate of the fingering to fracturing transition for transient self-assembled networks~\cite{Zhao1993,IgnesMullol1995,foyart_fingering_2013}. In this context, the interpretation in terms of a Deborah number defined as the ratio between a characteristic time scale of the experiment and a sample relaxation time seem relevant for some materials but largely fails for others~\cite{Zhao1993}. An open question here is understanding which time is pertinent for a specific experiment or process when it comes to multi-scale soft materials, whose properties are highly sensitive to various stimuli, including mechanical stress. In the context of drop impact experiments, we challenge here the too simplistic view based on one sample terminal relaxation time as the criterion for viscoelastic instabilities. In addition, most theoretical and experimental works on unstable flows of viscoelastic fluids, including elastic instabilities and elastic turbulence deal with solutions of polymer or living-polymers, and other types of viscoelastic networks have not been considered so far~\cite{pakdel_elastic_1996, fardin_elastic_2010, bonn_large_2011, bertola_experimental_2003, Gaillard, casanellas_stabilizing_2016, steinberg_elastic_2021}. By using different types of viscoelastic networks, we here assess also the possible roles of the network structure and of the structural susceptibility to a large deformation in a viscoelastic instability. For one class of network, the extreme shear following impact eventually weakens the junction points (inter-chain failure) whereas for the second class of network, it breaks the chains themselves (intra-chain failure).

We use an experimental set-up based on the free in air expansion of a viscoelastic sheet following the impact of a sample drop on a solid target of size comparable to that of the drop. We investigate two classes of viscoelastic fluids with different structures, for which we can tune the terminal (Maxwell) relaxation time and elastic modulus. The onset of viscoelastic instability for the different materials is quantitatively interpreted in term of a \textcolor{myc} {Deborah number based on the comparison between a characteristic experimental time and a sample intrinsic relaxation time.} We demonstrate that consistent results require to take into account a characteristic relaxation time following a nonlinear (fast) deformation of the sample. This work therefore highlights the role of the nonlinear processes to quantitatively rationalize a criterion for a viscoelastic instability.

The paper is organized as follows. We first present the viscoelastic samples and experimental set-up. We then sketch the phenomenology for the expansion of the viscoelastic sheets, and propose a quantitative criterion based on image analysis to define the onset of instability of the sheets leading eventually to holes. The experimental results are discussed and analysed using the quantitative criterion, and finally rationalized  in the light of the sample structure and related rheological properties.

\section{Materials and methods}
\subsection{Materials}

\begin{figure}[htb]
\centering
  \includegraphics[width=8.5 cm]{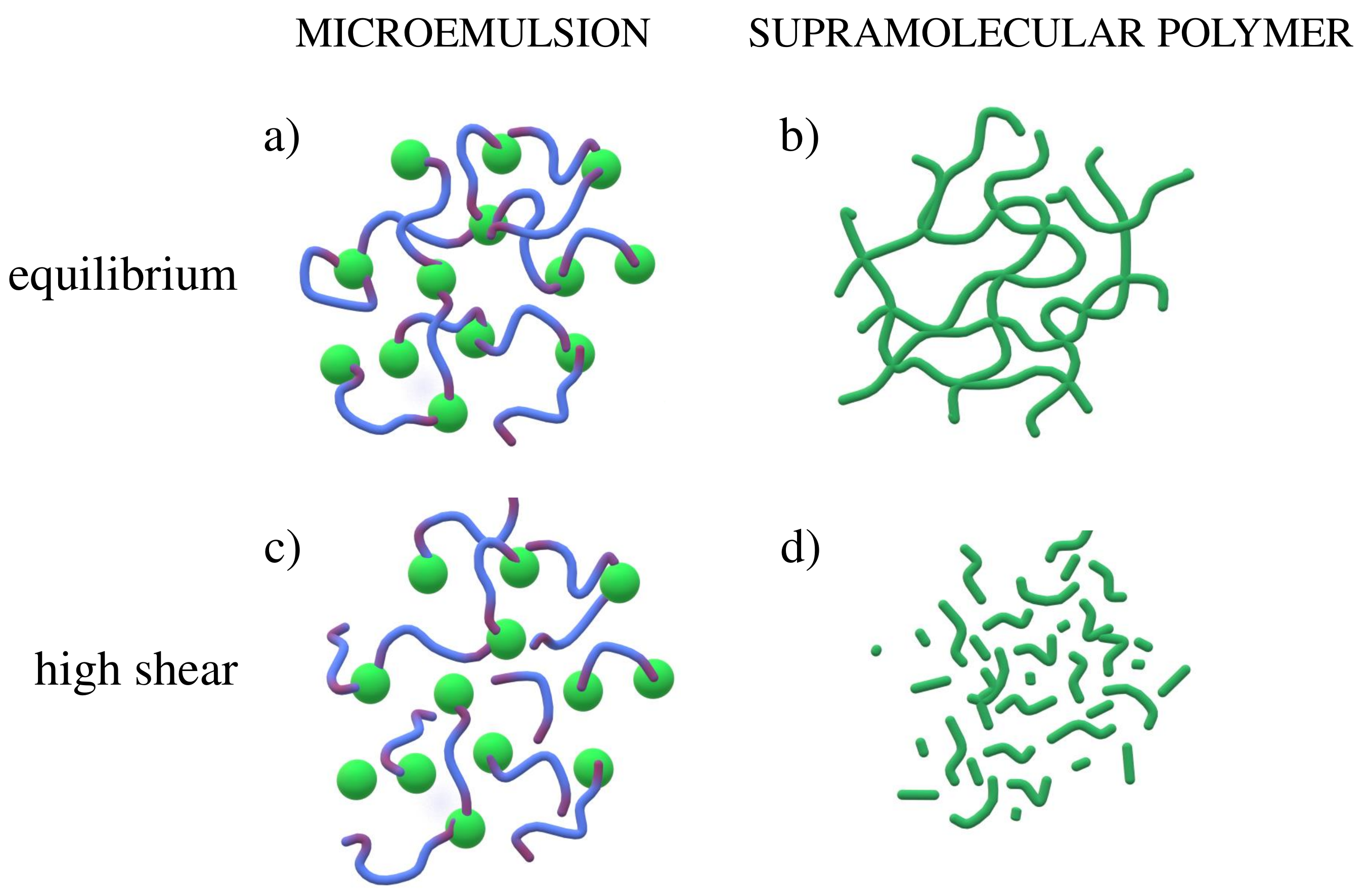}
 \caption{Schemes of the expected structures at rest (a,b) and under high shear (c,d) for the two types of transient networks, connected microemulsions (a,c) and semi-dilute solution of supramolecular polymer (b,d). }
 \label{fgr:Cartoons}
\end{figure}

We investigate two classes of transient networks, connected microemulsions and semi-dilute solutions of supramolecular polymers. Cartoons of the two classes of networks are provided in Figure~\ref{fgr:Cartoons}.

Connected microemulsions consist in surfactant-stabilized oil droplets~\cite{michel_percolation_2000, arora_interplay_2016}, dispersed in brine ($0.2$ M Nacl), and reversibly bridged by telechelic polymers. The oil (decane) droplets have a diameter of $6$ nm and are stabilized by a mixture of cetypyridinium chloride (CpCl) and octanol, with a molar ratio of octanol/CpCl of $0.65$. Telechelic polymers are triblock copolymers made of a long water-soluble polyethylene oxide chain of molecular weight $35$ kg/mol, end-grafted with short carboxylic chains with $n=12$ or $n=18$ carbons.
We fix the average number of telechelic polymer molecules per oil droplet, $r=2$, and vary the concentration of oil droplets $C$ between $0.5$  and $10$ g/L. \textcolor{myc}{A dye (erioglaucine, concentration $2.5$ g/L) is added to the samples with $n=18$. We mention that the addition of a dye has been used previously to measure the thickness of sheets during their expansion~\cite{lastakowski_bridging_2014,vernay_free_2015,wang_drop_2017,raux_spreading_2020}. Here we simply use the erioglaucine dye to enhance the contrast for the imaging of the sheets. Note moreover that with this dye, the light intensity does not follow the standard Beer-Lambert law~\cite{vernay_free_2015,wang_drop_2017}.}

The second class of viscoelastic fluids is semi-dilute solutions of supramolecular polymers in dodecane.  The supramolecular polymers are made of 2, 4-bis (2-ethylhexylureido) (abbreviated as EHUT) molecules, whose synthesis has been reported elsewhere~\cite{lortie_structural_2002}. Samples with EHUT concentration $C$ between $0.37$ g/L and $3$ g/L are prepared by adding EHUT powder, as synthesized, to the appropriate volume of dodecane and stirring the solution at a temperature of $80^{\circ}$C for $48$ hours. The final suspensions are transparent.  At room temperature and in the range of concentration investigated here, the EHUT molecules self-assemble by means of hydrogen bonding into tubes with a cross-section comprising about three EHUT molecules~\cite{bouteiller_thickness_2005}. The tubes are long enough to entangle and confer viscoelasticity to the suspensions~\cite{ducouret_rheological_2007,louhichi_humidity_2017}.

\begin{figure}[htb]
\centering
  \includegraphics[width=8.5 cm]{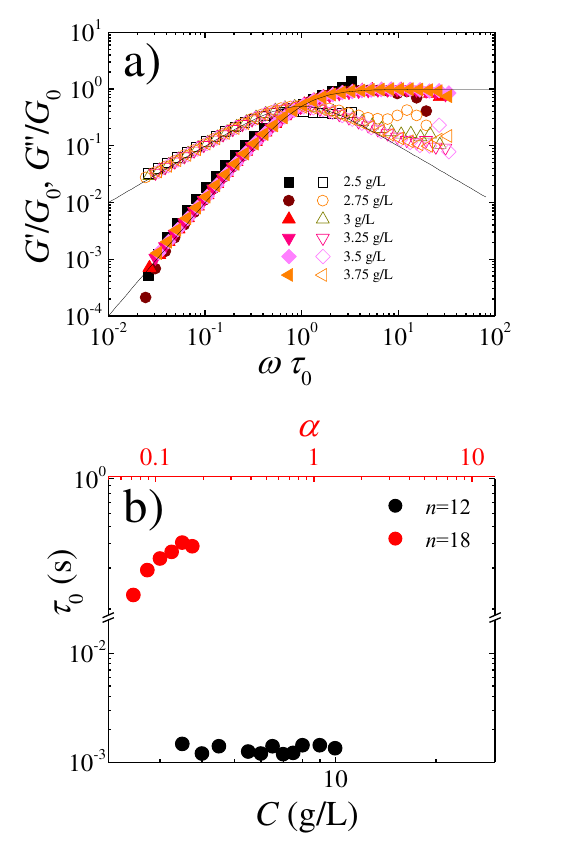}
 \caption{(a) Linear viscoelasticity in normalized units for transient networks comprising oil droplet reversibly linked by telechelic polymers with $n=18$ and various concentration $C$ as indicated in the caption. Symbols are data points (solid symbols correspond to $G'$ and empty symbols correspond to $G"$) and lines are the Maxwell-fluid model in normalized units. (b) Characteristic relaxation time as a function of $C$ (bottom $x$-axis) and as a function of the relative distance from the percolation threshold \textcolor{myc} {$C_{\rm{p}}$} (top $x$-axis) for microemulsion-based samples with $n=12$ \textcolor{myc}{($C_{\rm{p}}=2.2$ g/L)} and $n=18$ \textcolor{myc}{($C_{\rm{p}}=2$ g/L)}.}
 \label{fgr:ViscoelasticityMicroemulsion}
\end{figure}

\begin{figure}[htb]
\centering
  \includegraphics[width=8.5 cm]{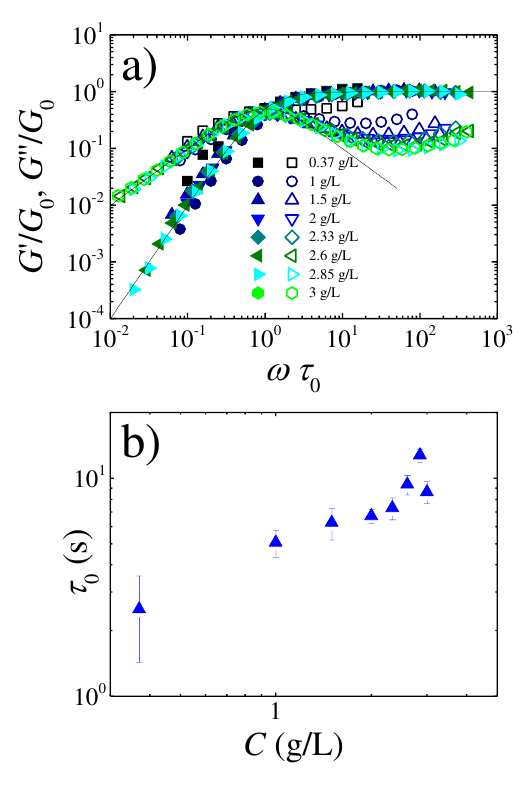}
 \caption{(a) Linear viscoelasticity in normalized units for transient networks semi-dilute solutions of self-assembled supramolecular polymers with various concentration $C$ as indicated in the caption. Symbols are data points (solid symbols correspond to $G'$ and empty symbols correspond to $G"$) and lines are the Maxwell-fluid model in normalized units. (b) Characteristic relaxation time as a function of $C$. }
 \label{fgr:ViscoelasticityEHUT}
\end{figure}

For both classes of networks, the linear viscoelastic behavior can be very well described in the low frequency window by a Maxwell fluid model, where the frequency-dependence storage, respectively loss, moduli read
$G'(\omega)=\frac{G_0 (\omega \tau_0)^2}{1+ (\omega \tau_0)^2}$, respectively $G"(\omega)=\frac{G_0 \omega \tau_0}{1+ (\omega \tau_0)^2}$. Here $\omega$ is the frequency, $G_0$ is the elastic plateau modulus and $\tau_0$ the characteristic relaxation time. The frequency dependence of the $G'$ and $G"$ together with the best fit to a Maxwell model is shown in Figure~\ref{fgr:ViscoelasticityMicroemulsion}
(a) for microemulsion-based samples with $n=18$ and in Figure~\ref{fgr:ViscoelasticityEHUT}(a) for supramolecular polymer-based samples, both at different concentrations. In the two graphs we have plotted the data in normalized units, $G'/G_0$ and $G"/G_0$ as a function of $\omega \times \tau_0$, where $G_0$ and $\tau_0$ are extracted for each sample from a fit with a Maxwell model. The collapse of the data acquired for different samples ensure a good agreement of the viscoelasticity with the Maxwell model \textcolor{myc} {at low frequency}.
\textcolor{myc}{For both classes of networks, deviations from the single-mode Maxwell model are observed at high frequency for the loss modulus, which exhibits a minimum at a characteristic frequency. This frequency is related to the characteristic time for bond-breaking and recombination process for supramolecular-based samples ~\cite{granek_stress_1992,cates_rheology_2006}, and is presumably linked to the Rouse dynamics of the polymer chains that link the oil droplets for the microemulsion-based samples.}

All supramolecular polymer solutions are viscoelastic. The characteristic relaxation time, $\tau_0$ increases from $2.5$ to $12.7$ s as $C$ increases from $0.37$ to $3$ g/L (Fig.~\ref{fgr:ViscoelasticityEHUT}(b)). By contrast, for microemulsion-based samples, one can determine a percolation threshold, $C_{\rm{p}}$, below which the samples are Newtonian liquids and above which viscoelasticity emerges ($C_{\rm{p}}=2.2$ g/L for $n=12$ and $C_{\rm{p}}=2$ g/L for $n=18$)~\cite{arora_interplay_2016}. \textcolor{myc}{Above the percolation threshold, the elastic modulus and relaxation time vary as a power law with $\alpha$, the relative distance from the percolation threshold $\alpha=\frac{C-C_{\rm{p}}}{C_{\rm{p}}}$}.

Above percolation, for samples with $n=12$, the characteristic relaxation time is systematically measured to be of the order of $1$ ms (independently of the oil droplet concentration), corresponding
roughly to the experimental limit of measurable relaxation time (Fig.~\ref{fgr:ViscoelasticityMicroemulsion}(b)). For samples with $n=18$, $\tau_0$ weakly increases with $C$ from $0.10$ to $0.34$ s (Fig.~\ref{fgr:ViscoelasticityMicroemulsion}(b)).

\subsection{Impact experiments}

Thin sheets freely expanding in air are produced by impacting a single drop of material on a solid target of size comparable to that of the drop~\cite{rozhkov_impact_2002, rozhkov_dynamics_2004, vernay_free_2015, arora_interplay_2016}. In brief, we use a hydrophilic cylindrical target of diameter $d_{\rm{t}}=6.5$ mm, slightly larger than the drop diameter. The size of the drops is dictated by the inner diameter of the syringe (here $4$ mm) and the equilibrium surface tension. For microemulsion-based, respectively, supramolecular polymer-based, samples, surface tension is $28$ mN/m, respectively $25$ mN/m, yielding drops of diameter $d_0=(3.6 \pm 0.1)$ mm, respectively $d_0=(3.9 \pm 0.2)$ mm. The drops are injected from a syringe pump operating at a flow rate $1$ ml/min, through the needle placed vertically above the target. Most experiments are performed with the drops falling from a distance $h=91.0$ cm, yielding a velocity at impact of $v_0 = \sqrt{2gh} \simeq 4$ m/s ($g=9.81$ $\rm{ms}^{-2}$ is the acceleration due to gravity), but additional experiments are conducted with $h$ in the range $(11-111)$ cm, yielding $v_0$ in the range $(1.47-4.67)$ m/s. Time series of images are recorded after the impact of the drop using a high speed camera Phantom V7.3 (detector size $800$ pixels $\times$ $600$ pixels), operated at $6700$ frames per second.

\section{Experimental results and discussion}

\subsection{Phenomenological behavior of the viscoelastic sheets}

\begin{figure}[h]
\centering
  \includegraphics[width=8.8 cm]{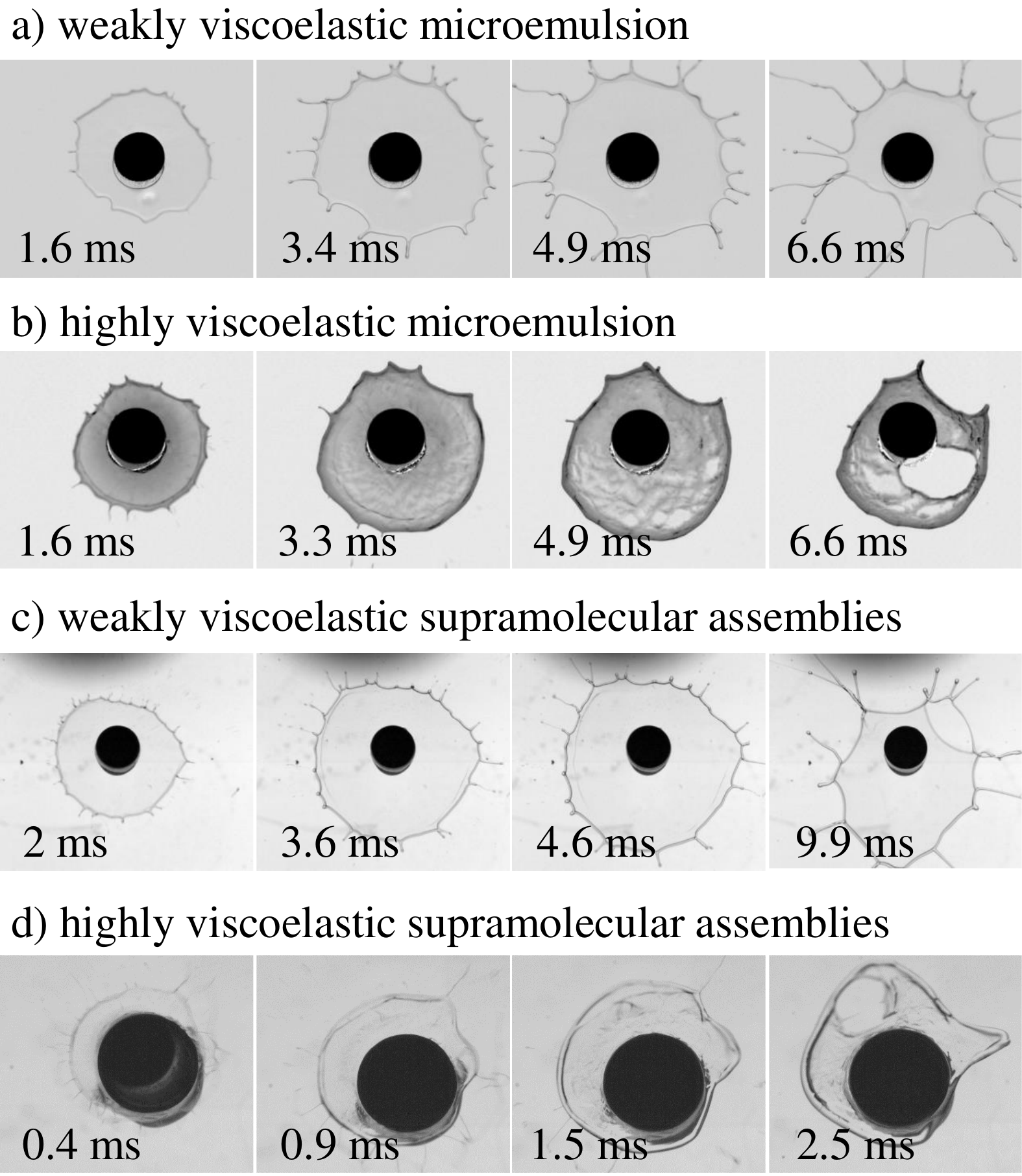}
 \caption{{Times series of sheet spreading and retraction for (a, b) viscoelastic microemulsion-based samples, with (a) $n=12$ and $C=3.5$ g/L ($\alpha=0.59$), and (b) with $n=18$ and $C=2.75$ g/L ($\alpha=0.375$), and (c, d) supramolecular polymer-based samples, with (c) $C=0.37$ g/L and (d) $C=3$ g/L. The maximum expansion of the sheet occurs at (a) $t_{\rm{max}}=4.5$ ms (b) $t_{\rm{max}}=4.2$ ms, (c) $t_{\rm{max}}=6.5$ ms and (d)  $t_{\rm{max}}=1.6$ ms. The scales are set by the target diameter, $d_{\rm{t}}=6.5$ mm (black disks).}}
 \label{fgr:images}
\end{figure}

\begin{figure}[h]
\centering
\includegraphics[width=8.5 cm]{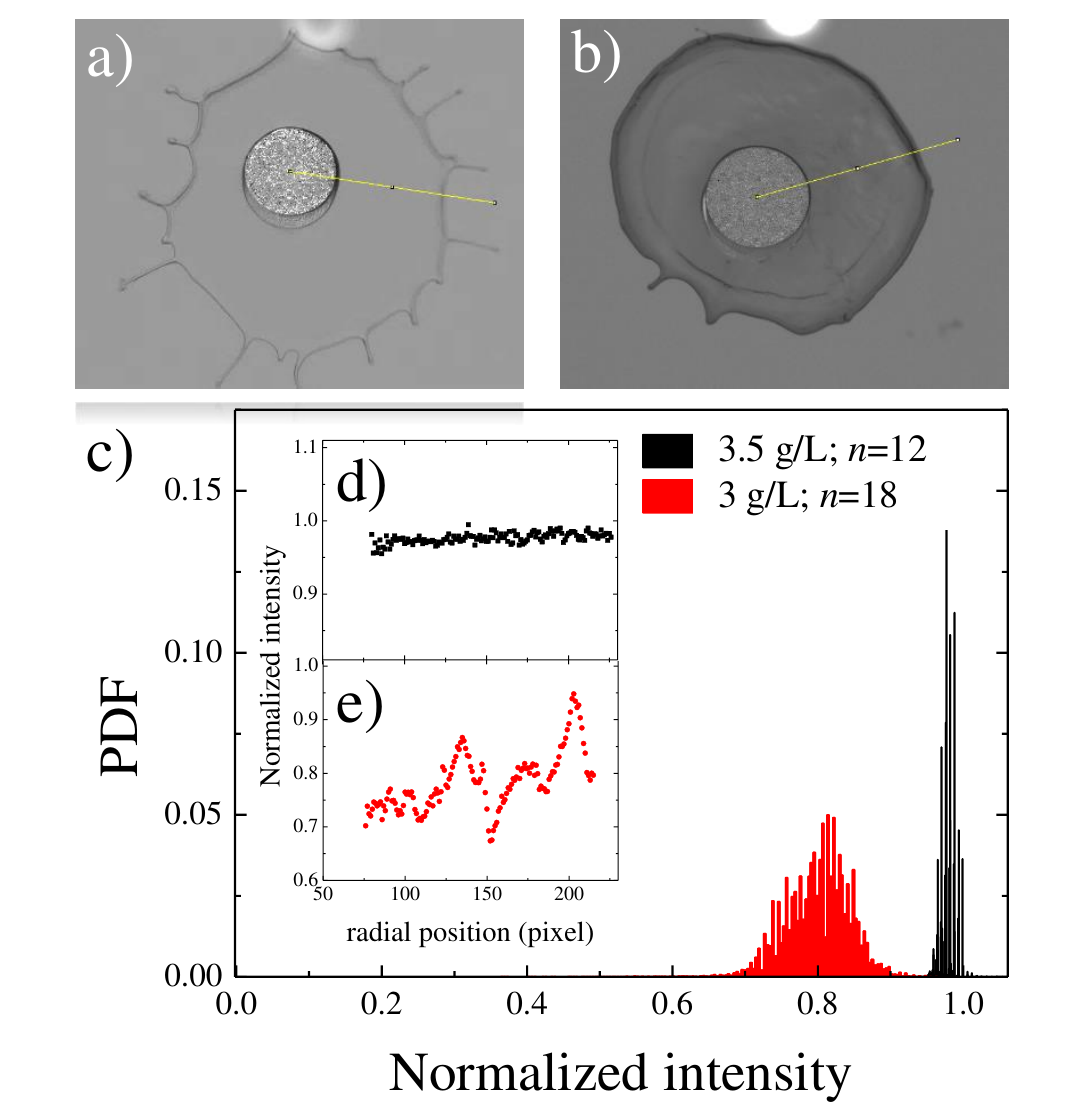}
 \caption{(a,b) Images taken at the maximal expansion for microemulsion-based samples, with $n=12$, $C=3.5$ g/L ($\alpha=1.5$) (a) and  $n=18$, $C=3$ g/L ($\alpha=0.33$) (b). The scale is set by the target size (central disk) of diameter $6.5$ mm. Images have been divided by an images taken before the drop impact. (d,e) Intensity profiles as a function of the radial distance from
the target center as measured along the yellow lines shown in (a, b). (c) Histogram of the intensity of the sheet for the two images shown in (a, b). Note that the overall lower value of the intensity for the sample with $n=18$ as compared to the sample with $n=12$ is because the sample with $n=18$ contains a dye.}
 \label{fgr:Profiles}
\end{figure}

We first describe qualitatively the phenomenology following the impact of a drop on a small solid target.
Images illustrating the different behaviors are shown in Figure~\ref{fgr:images} for drops falling from a fixed distance $h=91$ cm, yielding an impact velocity $v_0 \simeq 4$ m/s. Figure~\ref{fgr:images}(a) displays a time series of images showing the whole process for a weakly viscoelastic microemulsion-based sample prepared with telechelic polymers with $n=12$ ($C=3.5$ g/L, $\alpha=0.59$, and $\tau=1.5$ ms). The behavior is rather standard: After the drop impact, an apparently smooth sheet freely expands in air. The sheet is bounded by a thicker rim that destabilizes into ligaments, which subsequently disintegrate into drops. The sheet then retracts due to surface tension and eventually bulk elasticity. This behavior is very similar to the one observed for low viscosity Newtonian samples~\cite{rozhkov_impact_2002, rozhkov_dynamics_2004, villermaux_drop_2011, vernay_free_2015, arora_interplay_2016}. Following the impact on the target, the thickness of the sheet uniformly evolves with time and radial position. However, we find that the profile is nearly uniform all over the sheet at maximal expansion (see the profile in Fig.~\ref{fgr:Profiles}(d)), as previously measured also for a sheet made of pure water~\cite{vernay_free_2015}.

Although the overall picture, formation, expansion and then retraction of the sheet, is preserved, we observe a distinct behavior for a viscoelastic sample prepared with telechelic polymer with $n=18$ ($C=2.75$ g/L, $\alpha=0.375$, and $\tau=0.20$ s) (Fig.~\ref{fgr:images}(b)). First, the destabilization of the rim into filaments seems inhibited, presumably due to the high sample viscosity, in accordance with observation with highly viscous Newtonian liquids~\cite{arora_interplay_2016}. Second, more importantly and at odds  with observations for sheets produced with viscous Newtonian fluids, \textcolor{myc}{the brightness is found to be highly irregular in space and time, as revealed by the inhomogeneity in the grey level of the images of the sheets. Since the irregularities lead eventually to the opening of holes (Fig.~\ref{fgr:images}(b) and Fig.~\ref{fgr:Cracks}(a)), we believe that they originate mainly from variations of the thickness of the sheet, although we cannot exclude the contribution of localized inclinations of the sheet.} Interestingly, the two distinctive behaviors, highlighted for weak and strong viscoelastic microemulsion-based samples, can also been distinguished for the viscoelastic supramolecular polymer samples. Smooth and regular sheets are observed at low concentration ($C<2$ g/L, Fig.~\ref{fgr:images}(c)), whereas rough and irregular ones are observed at higher concentration ($C>2$ g/L, Fig.~\ref{fgr:images}(d)).
We mention that the maximum expansion of the sheets is also measured to strongly vary among samples. This effect has been rationalized in terms of rate-dependent bi-extensional viscosity for the supramolecular polymer-based samples and for semi-dilute solutions of water-soluble linear polymers~\cite{louhichi_biaxial_nodate, louhichi_competition_2021} and will not be commented further. In the following, we focus on the smoothness or roughness nature of the sheets, a parameter which has not be quantitatively considered so far, and we quantify this parameter at the maximal expansion of the sheets.

\subsection{Onset of instability}

The prominent feature of some viscoelastic sheets is the irregularity of their \textcolor{myc}{brightness}. Here, in order to investigate the irregular character of the thickness of the sheet, we propose to quantify the fluctuation in the brightness of the images of the sheets. Figure~\ref{fgr:Profiles} summarizes our experimental method, using the software Image J. To avoid artefact, images of the expanding sheet are first divided by a blank image taken just before the impact of the drop. As an illustration, we show in Figure~\ref{fgr:Profiles}(d,e) intensity profiles for two sheets, as obtained from the images shown in Figure~\ref{fgr:Profiles}(a,b). Overall, the intensity increases as the image is brighter and the thickness is lower. The profile for the microemulsion-based viscoelastic sample with $n=12$ is smooth and almost flat, in sharp contrast to the profile for the sample $n=18$. The fluctuations in the profiles are quantified by computing the histogram  of the intensity for all pixels belonging to the sheet (excluding the thicker rim and the part, not detected, on the surface area of the target). One defines the relative standard deviation (RSD) as the ratio between the standard deviation, $\sigma$, and the average of the pixel values $RSD= \frac{\sigma}{I_{\rm{mean}}}$. Here $\sigma$ and $I_{\rm{mean}}$ are extracted from the histogram.
We moreover normalize the RSD by the constant low values measured below percolation for microemulsion-based samples and at low concentrations for supramolecular-polymer based samples, $RSD_0$. We thus define the relative contrast of a sheet as $\chi=\frac{RSD_{\rm{sample}}}{RSD_0}$. A relative contrast close to $1$ is the signature of smooth sheets and reflects also the fact that the thickness of the sheets varies only very weakly with the distance form the center of the sheet at maximal expansion. In the following, we compute and show the relative contrast at the maximal expansion of the sheet, \textcolor{myc}{and we use this parameter as a marker for the onset of the instability of the sheets. Quantifying how heterogeneities of the sheets evolve with the distance from the onset of instability is beyond the scope of this work.}

\begin{figure}[tb]
\centering
  \includegraphics[width=8.5 cm]{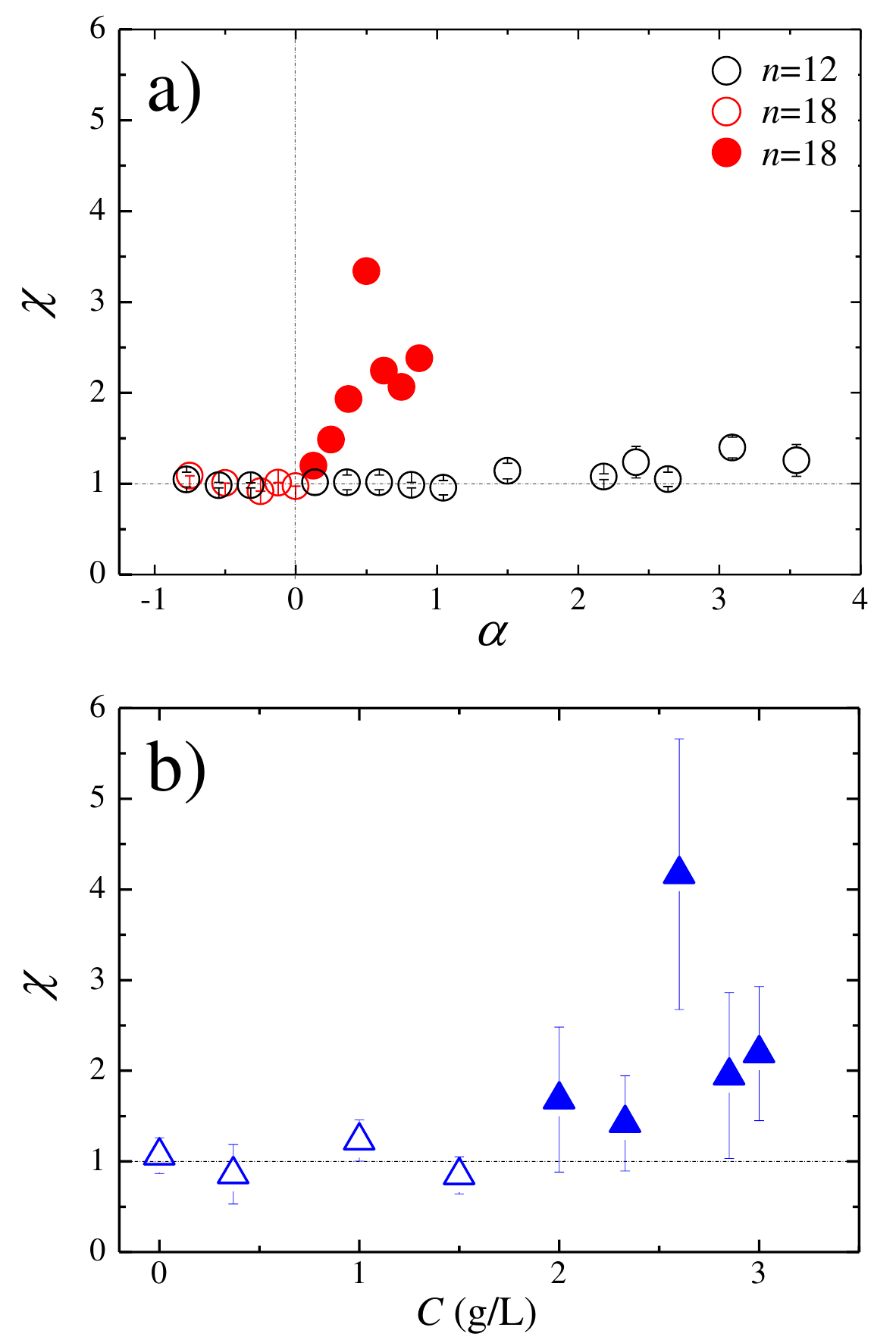}
 \caption{Relative contrast of viscous and viscoelastic sheets, taken at their maximal expansion, as a function of (a) the relative departure for percolation threshold for microemulsion-based samples prepared with telechelic polymers with $n=12$ and $n=18$, and (b) concentration for supramolecular polymer-based samples. For microemulsion-based samples, the error bars are smaller than the symbol size.}
 \label{fgr:contrast}
\end{figure}

\begin{figure}[htb]
\centering
\includegraphics[width=8.5 cm]{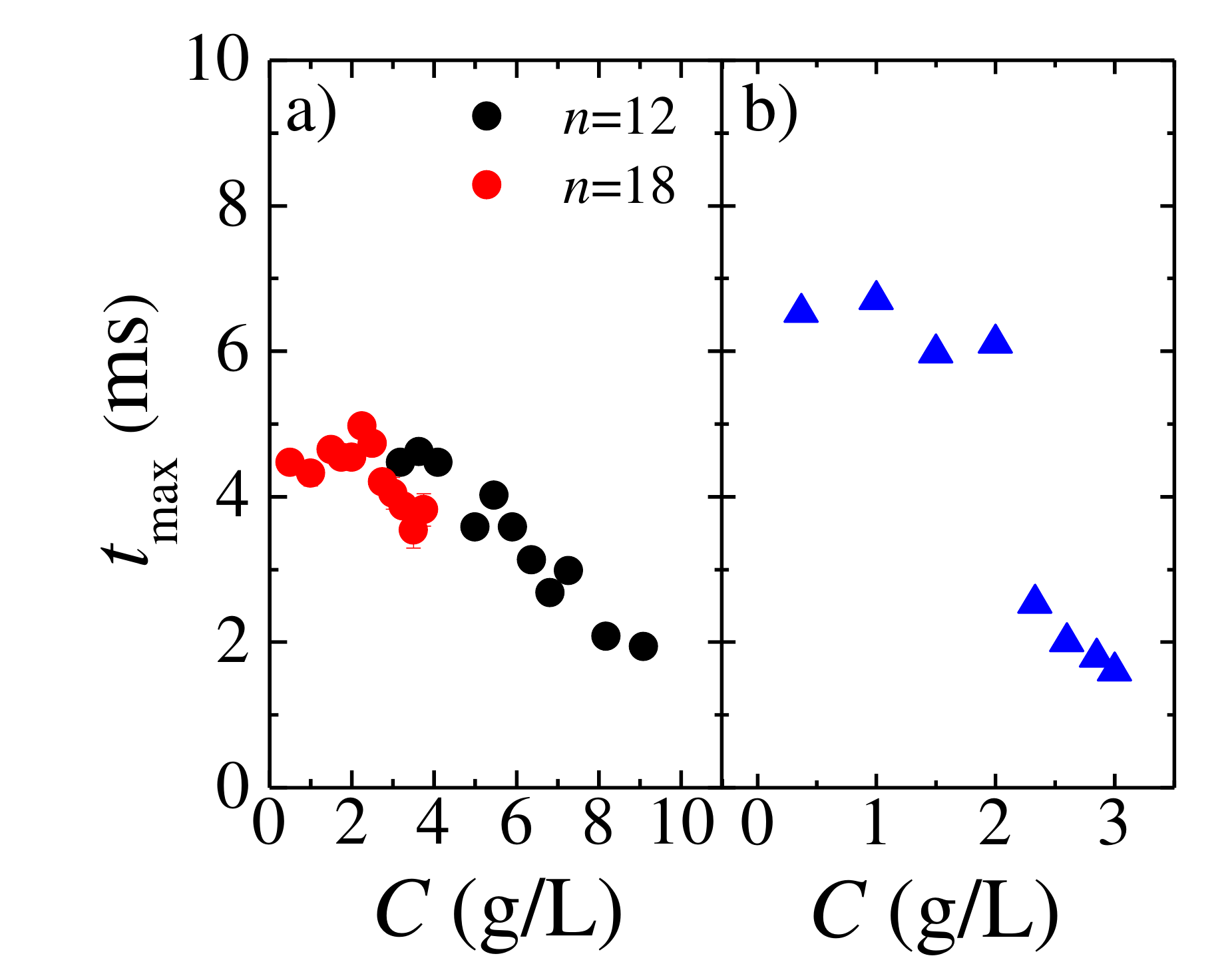}
 \caption{\textcolor{myc} {Duration of the process, from drop impact to maximal expansion of the sheet}, as a function of concentration for (a)  microemulsion-based samples, (b)  supramolecular polymer-based samples.}
 \label{fgr:tMax}
\end{figure}

\begin{figure}[hbt]
\centering
  \includegraphics[width=8.5 cm]{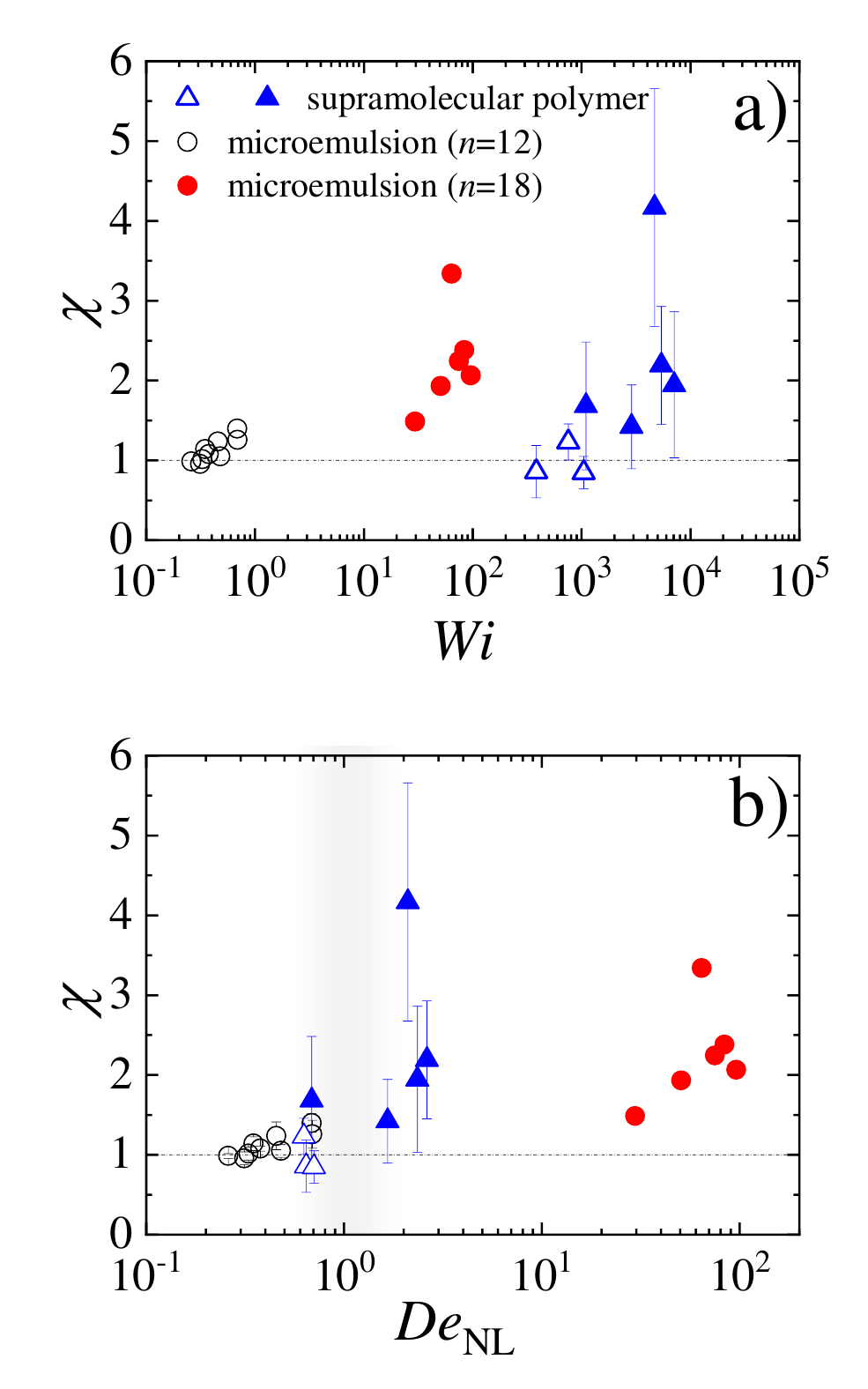}
\caption{Relative contrast as a function of (a) an apparent Deborah number, (b) a non-linear apparent Deborah number (see text) for the three sets of samples. Full, respectively empty, symbols indicate unstable, respectively stable, sheets. For microemulsion-based samples, the error bars are smaller than the symbol size. The shaded area in (b) indicates the transition from stable to unstable sheets.}
 \label{fgr:De}
\end{figure}

The relative contrast, $\chi$, is plotted as a function of the relative departure from the percolation threshold, $\alpha$, for microemulsion-based samples (Fig.~\ref{fgr:contrast}(a)) and as a function of EHUT concentration $C$ for supramolecular polymer-based samples (Fig.~\ref{fgr:contrast}(b)). In accordance with the qualitative observations (Fig.~\ref{fgr:images}), we find that the sheets prepared with microemulsion comprising telechelic polymers with $n=12$ are all smooth ($\chi \approx 1$). Similarly, the sheets made with viscous microemulsion-based samples (below percolation, $\alpha <0$) prepared with telechelic polymers with $n=18$ are as smooth as sheets prepared with pure solvent and with microemulsions with $n=12$ ($\chi\approx1$). By contrast, the viscoelastic sheets ($\alpha > 0$) are more irregular ($\chi > 1$). For viscoelastic sheets, $\chi$ reaches numerical values of up to $3.5$, with a maximum at an intermediary $\alpha$. Interestingly, and in contrast with the findings for microemulsion-based samples with $n=18$, all viscoelastic supramolecular polymer-based samples do not yield sheets with heterogeneous \textcolor{myc}{brightness}, but only the more concentrated ones ($C \gtrsim 2$ g/L). Above this threshold concentration, similarly to our findings for emulsion-based samples, we find that $\chi$ exhibits a non-monotonic evolution with $C$ with a maximum around $4$ for $C=2.6$ g/L.
In the following, we refer to unstable sheets whenever $\chi >1$.

\subsection{Analysis in terms of a \textcolor{myc}{relevant Deborah} number}

Here we provide a rationalization for the emergence of sheet instabilities, leading eventually to hole opening, based on the link between the rheological relaxation times of the samples and the relevant timescales of the drop impact experiments.
As discussed by us recently~\cite{louhichi_biaxial_nodate,louhichi_competition_2021}, during the process of sheet formation and expansion, the sample experiences a combination of shear (on the target) and biaxial deformation when the sheet expands. \textcolor{myc} {Based on our recent works~\cite{louhichi_biaxial_nodate,louhichi_competition_2021,charles_viscoelasticity_2021}, we know that the behavior of the sheets is essentially governed by their free biaxial expansion in air.}


The microemulsion-based samples below percolation (Figs.~\ref{fgr:images} and~\ref{fgr:contrast}) are Newtonian liquids. They are found to produce smooth and regular sheets. By analogy with these findings, we expect intuitively that smooth and regular sheets will be produced whenever viscoelastic samples are expected to behave as viscous ones. This should be the case whenever the relevant experimental time is longer than the sample relaxation time.  Hence, we propose, as a first-order simple description, that the relevant parameter controlling whether a viscoelastic sheet will tear or remain smooth and regular as a viscous sample is a \textcolor{myc} {Deborah} number defined as \textcolor{myc} {$De=\tau_0 /t_{\rm{max}}$, with $t_{\rm{max}}$, the time at which the sheet reaches its maximal expansion,} and $\tau_0$ the terminal relaxation time (i.e. the characteristic time of the Maxwell element) determined through linear viscoelastic measurements (Figs.~\ref{fgr:ViscoelasticityMicroemulsion} and~\ref{fgr:ViscoelasticityEHUT}). \textcolor{myc} {The evolution of $t_{\rm{max}}$ with sample concentration is shown in Figure~\ref{fgr:tMax}. We find overall comparable numerical values for $t_{\rm{max}}$ (of the order of a few ms) for all samples with a typical time that decreases as $C$ increases. Note that, operationally, $1/t_{\rm{max}}$ can be regarded as an apparent expansion rate of the sheet during its expansion.}

We first test this hypothesis with the microemulsion-based samples. They consist of spherical oil droplets connected by telechelic polymers and forming a percolated network (Fig.\ref{fgr:Cartoons}), a robust structure, especially with respect to mechanical deformation. (This point will be discussed more in detail below.) We plot in Figure~\ref{fgr:De}(a), for all viscoelastic microemulsion-based samples, the relative contrast as a function of \textcolor{myc} {$De$}. Consistent results are obtained: for all samples prepared with a telechelic polymer with $n=12$,  \textcolor{myc} {$De<1$} and the sheets prepared with these samples do not exhibit instabilities. By contrast for the samples prepared with a telechelic polymer with $n=18$, \textcolor{myc} {$De>1$}, and the sheets are found to be unstable. These findings confirm the relevance of the simple approach described above.

For supramolecular polymer-based samples, \textcolor{myc} {$De \gg 1$} ( \textcolor{myc} {$De$} ranges between $\sim 400$ and $\sim 6000$) because of much longer relaxation times than for microemulsion-based samples for comparable \textcolor{myc} {experimental time}. However, despite these very large values, we find that the less concentrated samples (with  \textcolor{myc} {$De$} up to $\sim 1000$) produce smooth and regular sheets, in contrast to the results of microemulsion-based samples and associated expectations, thus challenging the simple physical picture presented above (Fig.~\ref{fgr:De}(a)).

\begin{figure}[h]
\centering
\includegraphics[width=8.5 cm]{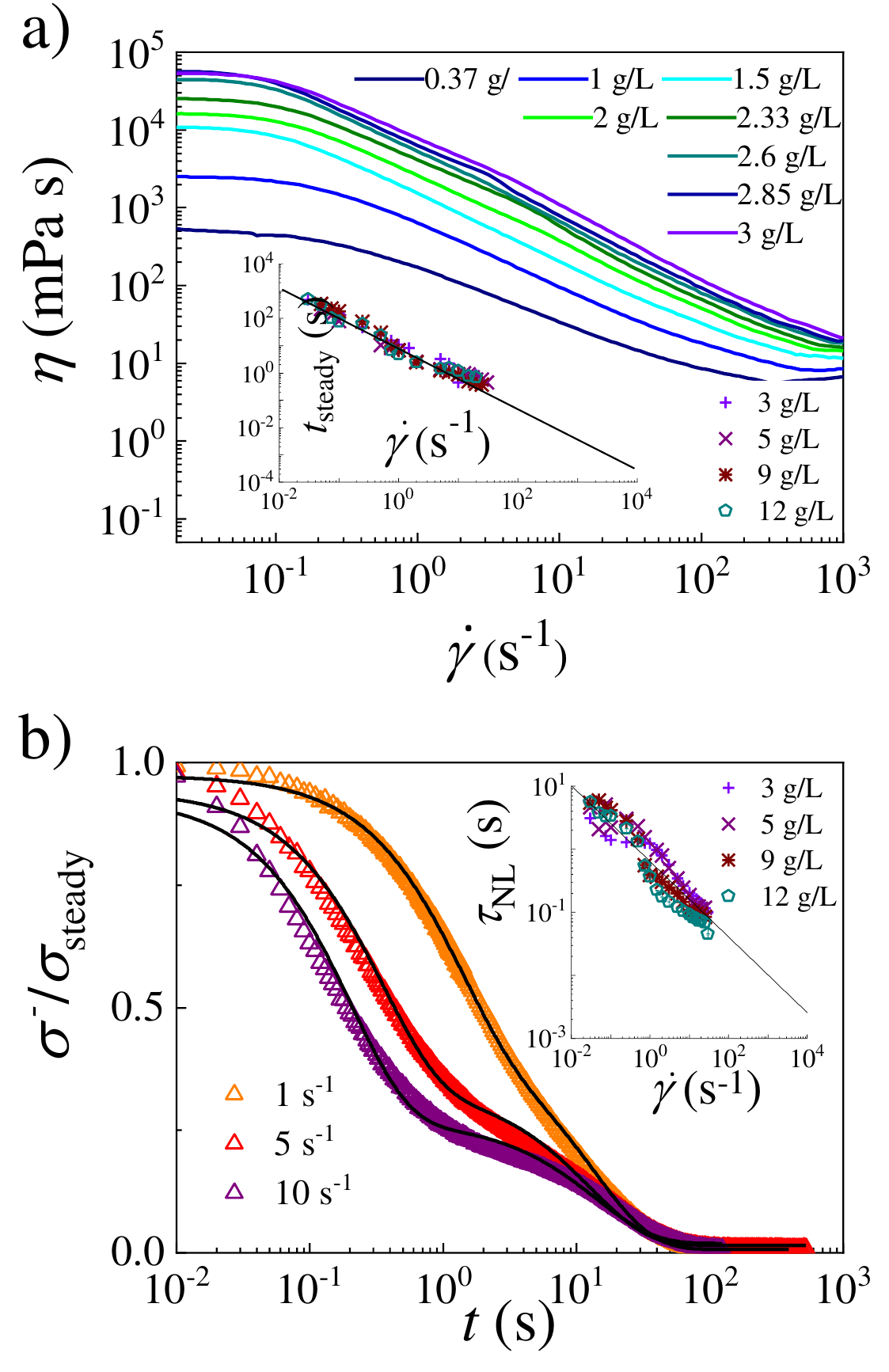}
 \caption{(a) Steady state viscosity as a function of shear rates for supramolecular polymer-based samples with different concentrations as indicated in the legend. (Inset) Time needed to reach the steady state as a function of the  shear-rate imposed for various EHUT concentration as indicated. (b) Stress relaxation following a steady shear flow at different rates as indicated in the caption, for a supramolecular polymer-based sample with $C=3$ g/L. The stress is normalized by the steady shear stress reached during steady shear flow. (Inset) Non-linear characteristic relaxation time as a function of the shear-rate (see text). In the two insets, symbols are experimental data points and lines are power law fits of the experimental data.}
 \label{fgr:NonLinear}
\end{figure}

\begin{figure}[h]
\centering
\includegraphics[width=8.5 cm]{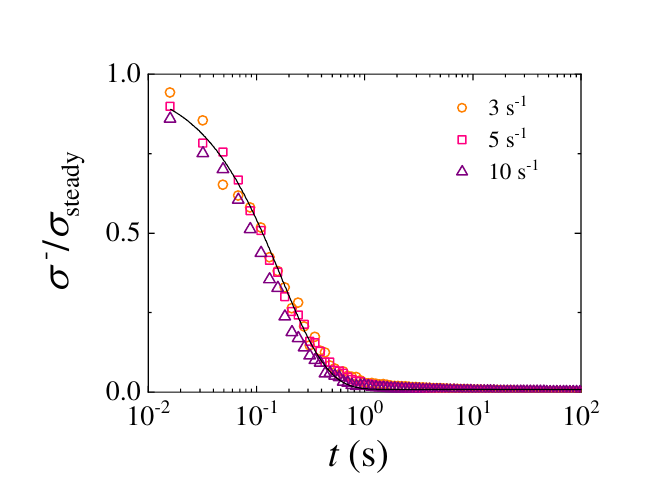}
 \caption{Stress relaxation following a steady shear flow at different rates as indicated in the caption, for a microemulsion-based sample with $n=18$, and $C=3$ g/L. The stress is normalized by the steady shear stress reached during steady shear flow.}
 \label{fgr:NonLinearMicroEmulsion}
\end{figure}

To rationalize this apparent contradiction, we should recall that the sheets experience very large shear during their spreading on the solid surface of the target. In Ref.~\cite{louhichi_competition_2021}, we provide a relationship for the mean shear rate experienced by the sheet on a solid target, as averaged over the whole duration of the sheet expansion: $\dot{\gamma}_{\rm{av}}\approx \frac{v_0 d_{\rm{t}} d_{\rm{max}}}{4d_0^3}$. Here $v_0\approx 4$ m/s is the impact velocity, $d_{\rm{t}}=6.5$ mm is the target diameter, $d_0=(3.6-3.9)$ mm is the initial drop diameter, and $d_{\rm{max}}$ is the sheet diameter at maximal expansion, which weakly varies from sample to sample. For supramolecular-based samples, we estimate  $\dot{\gamma}_{\rm{av}}\approx 2000$ $\rm{s}^{-1}$.

As shown in Figure~\ref{fgr:NonLinear}(a) that displays the steady state shear viscosity as a function of the imposed shear rate, the supramolecular polymer-based samples are strongly shear-thinning. For a shear rate $\dot{\gamma}_{\rm{av}}$ we evaluate, thanks to a fit of the data with a Cross model~\cite{cross}~\footnote{The cross model reads $\eta=\eta_{\infty} + \frac{\eta_0-\eta_{\infty}}{1=(k\dot{\gamma}^n}$ with $\eta_{\infty}=1.5$ mPas, the solvent viscosity, $\eta_0$ the zero shear viscosity and $k$ and $n$ two fit parameters} that the sample viscosity in the steady state is closed to that of the solvent ($1.5$ mPas), for all concentrations. We indeed find that the viscosity ranges between $2$ and $7$ mPas. In addition, the time needed to reach the steady state viscosity, $t_{\rm{steady}}$, for shear rates in the range ($0.03-20$) $\rm{s}^{-1}$ is measured to be roughly inversely proportional to the imposed shear rate (inset Fig.~\ref{fgr:NonLinear}(a)), \textcolor{myc}{implying a total shear deformation larger than ca. $10$ is reached; this value is comparable to the effective deformation estimated during the sheet expansion}~\footnote{\textcolor{myc}{The typical shear deformation $(d_{\rm{max}} - d_0)/e$, with $e$ the typical thickness of the sheet (evaluated from volume conservation with the sheet viewed as a pancake of uniform thickness $e$), decreases from $50$ to $10$ as $C$ increases.}}
. As an additional check, by extrapolation of the data of the inset of Fig.~\ref{fgr:NonLinear}(a)) to a shear rate $\dot{\gamma}_{\rm{av}}$, we expect $t_{\rm{steady}}$ of the order of $1$ ms. This time is significantly shorter than the typical duration time of the expansion process. Therefore one can safely consider that the steady state viscosity is attained during the expansion process. Hence, when considering a sample relaxation time, one has to consider the relaxation time of a sample after it has been sheared at a prescribed shear rate. This time is evaluated experimentally from the stress relaxation upon cessation of shear flow~\cite{AmeurThesis, hendricks_nonmonotonic_2019}. The time evolution of the stress is shown in Figure~\ref{fgr:NonLinear}(b) for a sample with concentration $C=3$ g/L and three shear rates ($1$, $5$ and $10$ $\rm{s}^{-1}$). Time is set to $0$  when the steady shear flow is stopped and stresses are normalized by their value at time $0$, corresponding to  the steady state value of the stress during shear. A two-step relaxation is detected, with a shear-dependent first decay and a nearly shear-independent second decay.  Stress relaxation is fitted with a two exponential decay function (lines in Fig.~\ref{fgr:NonLinear}(b)), from which the characteristic time of the fast decay, $\tau_{\rm{NL}}$, is derived. Data for $\tau_{\rm{NL}}$ are reported in the inset of Figure~\ref{fgr:NonLinear}(b) for samples with varying concentrations $C$ between $3$ and $12$ g/L, and for shear rates spanning three orders of magnitude (from $0.03$ to $30$ $\rm{s}^{-1}$). We find that $\tau_{\rm{NL}}$ is nearly concentration independent and decays as a power law with the imposed shear rate with an exponent of the order of $-0.6$. We first discuss the physical meaning of the two relaxation times. For supramolecular polymer-based samples, at large shear rates, the steady state viscosity of the sample is very closed to that of the solvent, presumably because of the breaking of the intramolecular H-bonds that hold the EHUT molecules together to form long polymers~\cite{hendricks_nonmonotonic_2019} (Fig.~\ref{fgr:Cartoons}(b,d)). \textcolor{myc} {Accordingly, we expect the fastest stress relaxation process following the cessation of a high shear to be linked to the relaxation of short chains, which relax faster, contributing eventually to the formation of more longer chains. This process occurs at short length scales. Therefore, in the range of concentrations examined, the effect of concentration is expected to be weak, as observed experimentally. By contrast, the long relaxation time is linked to the stress-relaxation by reptation-like mechanism of the long supramolecular polymers in a semi-dilute regime. It is therefore shear-rate independent, but concentration dependent, as observed experimentally, and quantitatively comparable to the relaxation time measured in the linear regime (Fig.~\ref{fgr:ViscoelasticityEHUT}b)~\cite{AmeurThesis}}.
For the impact experiments, which involve fast processes, we argue that the relevant relaxation is the one associated with the fastest decay, $\tau_{\rm{NL}}$, which is evaluated from the extrapolation of the experimental data of Figure~\ref{fgr:NonLinear}(c) up to the relevant shear rate, $\dot{\gamma}_{\rm{av}}$.   This relaxation time ($\tau_{\rm{NL}}\simeq 4$ ms) is associated to non-linear processes and is much shorter than the linear equilibrium characteristic time by several orders of magnitude. Therefore the  \textcolor{myc} {relevant Deborah} number associated to a non-linear process defined as \textcolor{myc} { $De_{\rm{NL}} \simeq \tau_{\rm{NL}} /t_{\rm{max}} \ll De$ }. Using this \textcolor{myc} {Deborah} number, we find now consistent results for the three sets of data: all samples with \textcolor{myc} {$De_{\rm{NL}}>1$} exhibit unstable sheets yielding eventually to holes whereas the ones with  \textcolor{myc} {$De_{\rm{NL}}<1$}  lead to stable and smooth sheets (Fig.~\ref{fgr:De}(b)).

In summary, our experimental data (Fig.~\ref{fgr:De}(b)) suggest that consistent results are obtained when considering a linear relaxation time for the microemulsion-based samples but a non-linear one, associated to fast relaxation processes following a high shear of the supramolecular polymer-based samples. This seems counterintuitive and deserves some discussion regarding the physical process at play, for the two systems, based on their distinctive structural properties. As discussed above for supramolecular-based polymers, the relevant timescale to rationalize the fate of the expanding viscoelastic sheet is related to the process of recombination of broken links. On the other hand, microemulsion-based samples consist of surfactant-stabilized spherical oil droplets reversibly linked by telechelic polymers. They can be regarded as consisting of beads (the oil droplets) connected by springs (the telechelic polymers). Mechanical deformations, and in particular large shear fields, leave the beads intact but eventually decrease the average lifetime of the springs, i.e. the average residence time of the hydrophobic stickers of the telechelic copolymers in the oil droplets (Fig.~\ref{fgr:Cartoons}(a,c)). Upon cessation of shear, the physical process at the origin of stress relaxation is therefore related to the reorganization of the telechelic polymers among the droplets~\cite{Tanaka1992}. This process is similar to the process at the origin of the stress relaxation in the linear regime, for a sample at equilibrium. Hence, we expect the relaxation time following a high shear to be equal to the linear relaxation time. To test this, we have performed stress relaxation after cessation of shear for a sample with $C=3$ g/L and $n=18$. Interestingly, the behavior is strikingly different from the one measured with the supramolecular polymer-based sample (Fig.~\ref{fgr:NonLinear}). For the microemulsion-based sample, the stress relaxation curves after cessation of shear (for shear rate of $3$, $5$ and $10$ $\rm{s}^{-1}$) nicely superimpose, and display a monoexponential decay (line in Fig.~\ref{fgr:NonLinearMicroEmulsion}) yielding a characteristic time $0.16$ s, very close to the one measured in the linear regime ($\tau_0=0.1$ s). Thus, for microemulsion-based samples the relevant relaxation time here is comparable to the linear one, whereas for supramolecular polymer-based one, it is $3$ to $4$ orders of magnitude shorter than the linear one. We believe this difference is directly correlated to the different structures and flow mechanisms of the two classes of transient networks.

\subsection{Dynamics of holes opening}

\begin{figure}[tb]
\centering
\includegraphics[width=8.5 cm]{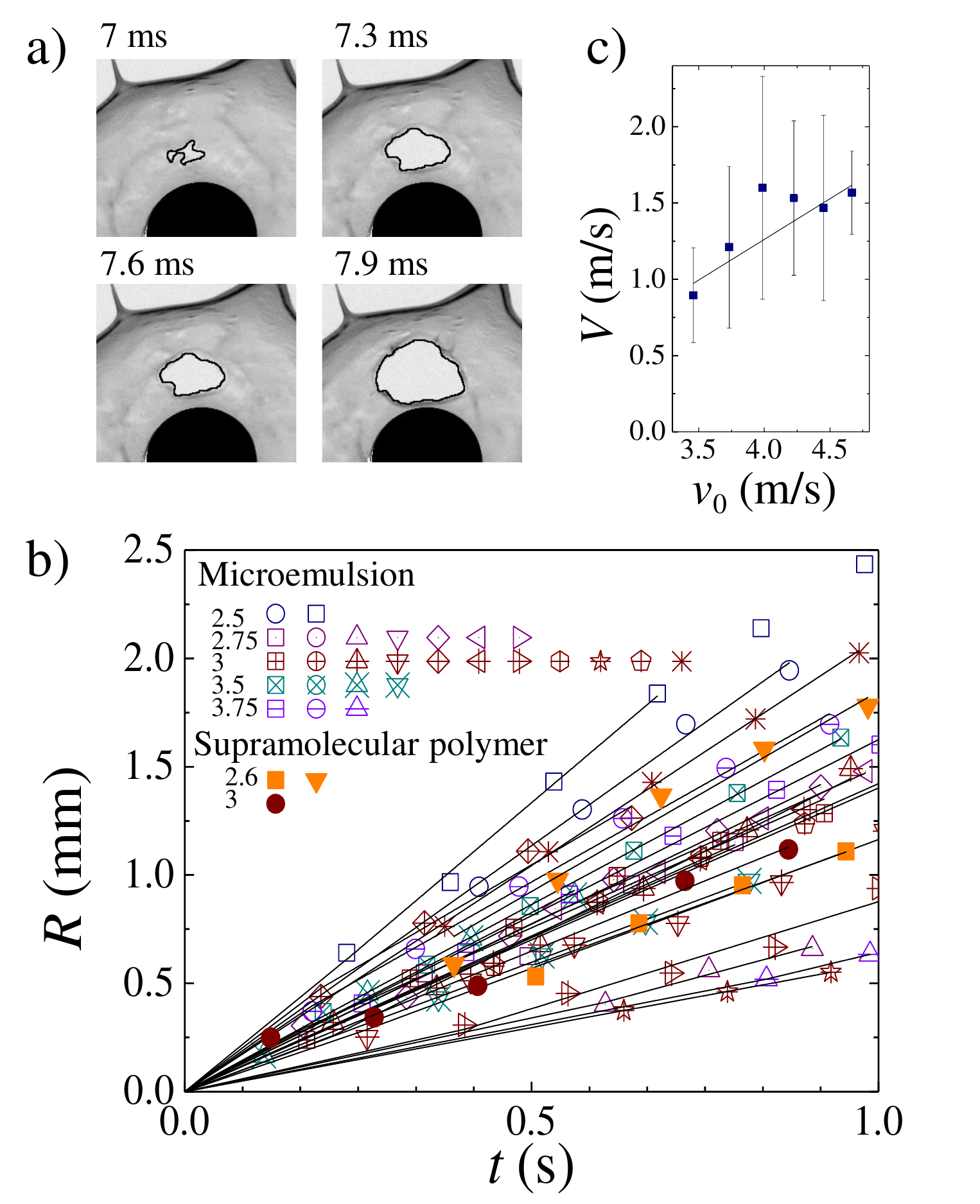}
 \caption{(a) Time series showing the expansion of a hole. The sample is a microemulsion linked with telechelic polymer with $n=18$ and $C=2.5$ g/L ($\alpha=0.25$). (b) Time evolution of the effective radius of holes, for all observed holes for all samples, for a fixed fall height $h=91$ cm. Time $0$ is here defined as the time at which each hole first appears. Symbols are sorted based on their fillings and colors, and correspond to different samples. Different shapes correspond to different holes for a given sample. (c) Average hole expansion velocity as a function of the impact velocity of the drops, for a microemulsion-based sample with $n=18$ and $C=2.75$ g/L ($\alpha=0.375$).}
 \label{fgr:Cracks}
\end{figure}

As described above, the \textcolor{myc}{instability} of the sheet may yield ultimately a hole, defined as a local opening of the sheet.
We show in Figure~\ref{fgr:Cracks} the evolution of the effective radius of a hole, $R=\sqrt{S/\pi}$ (with $S$ the surface area of the hole) with time $t$, where $t=0$ is the time at which each hole nucleates, obtained by the extrapolation of the data for the time evolution of $S$. All data acquired for microemulsion-based samples (with $n=18$), for different holes for a given sample concentration and for different sample concentration, and for supramolecular-polymer based samples for two concentration, display consistent behavior: the effective radius of holes varies linearly with time, allowing thus the determination of a hole opening velocity, $V$.  The velocity varies between $0.7$ and $2.5$ m/s, without any clear trend for a dependence with the sample concentration or the type of networks. A statistical analysis of $30$ data sets yields $V \approx 1.5 \pm 0.5$ m/s.

This characteristic velocity can be compared to the velocity of cracks that may occur in elastic materials. One can evaluate the transverse speed of sound for an elastic medium $V_s$ as $V_s=\sqrt{\frac{G_0}{\rho}}$. For the microemulsion-based samples, the elastic modulus varies between $4.4$ and $34$ Pa, yielding $V_s$ in the range $0.066$ and $0.18$ m/s. On the other hand, the elastic modulus of the supramolecular polymer solutions at concentration $C=2.6$ g/L (resp. $C=3$ g/L) is equal to $5.8$ Pa (resp. $8.5$ Pa) yielding $V_s$ in the range $(0.09-0.11)$ m/s. Hence, the measured velocity is larger by at least one order of magnitude than the speed of sound, which is the maximum velocity at which a crack can propagate in an elastic material at rest, ruling out such physical interpretation. \textcolor{myc} {We find that the measured velocity is directly correlated to the impact velocity of the drop, as inferred from experiments with drops of microemulsion-based samples that hit the target from different heights (Fig.~\ref{fgr:Cracks}(c)). Because of the irregular contours of the holes, we believe that the process at play here very likely differs from the much-studied hole opening phenomena for a soap film ~\cite{Taylor1959,Culick1960} or during the expansion following impact of an oil-in-water dilute emulsion~\cite{vernay_bursting_2015}, which are uniquely governed by surface tension and lead to circular holes.Hence, our observations suggest that the velocity of hole opening is not uniquely governed by the intrinsic properties of the viscoelastic materials, as in soap films.}

\section{Conclusions}

We have used impact experiments to investigate the behavior of viscoelastic fluids under an extreme deformation. We have studied two classes of transient networks: entangled solutions of supramolecular polymers made of EHUT molecules held together by hydrogen bonds (supramolecular polymer-based samples) and surfactant-stabilized spherical oil droplets of nanometer size suspended in water and reversibly linked together by telechelic polymers (microemulsion-based samples). For both systems, a series of samples have been investigated where the concentration in EHUT, for supramolecular polymer-based samples, and in oil droplets, for microemulsion-based samples, have been varied. Both classes of viscoelastic samples exhibit very similar linear, Maxwell-like, viscoelastic properties, but contrasting nonlinear properties due to marked differences in the structure of the transient networks. We have evidenced an instability in the viscoelastic sheets that form after a drop hits a small solid target. In all cases, after the drop impact on the small target, a sheet freely expands in air and then retracts, as observed for Newtonian fluids. By contrast, distinct features have been identified regarding the integrity of the sheet for the viscoelastic samples investigated in this study. Whether the sheet thickness evolves smoothly in space and time or display strong irregularities, also referred to as instabilities, leading eventually to the formation of holes, has been found to be specific to a subclass of the viscoelastic samples. We note here that the features of the sheets is reminiscent of the transition observed above a critical velocity in the texture of the sample-air interface, from smooth to rough, when a solid ball impacts a bath of viscoelastic fluid~\cite{akers_impact_2006}.

To rationalize our experimental findings, we have first provided a simple image analysis to determine the onset of instabilities, based on the quantification of the fluctuations of the pixel intensity, which are directly correlated to the fluctuations of the thickness of the sheet. We note that hole nucleation in liquid sheets has also been analyzed as originating from fluctuations of the thickness of viscous sheets, in a different context though (impulsive acceleration of a drop caused by a laser impact)~\cite{klein_drop_2020}. We have then demonstrated that this onset of instabilities is directly correlated to \textcolor{myc} {Deborah} number, defined as \textcolor{myc} {the ratio of a sample characteristic relaxation time over a typical experimental time}. Consistent data are obtained only if the relevant relaxation time is considered. This time, which we have measured for the two classes of samples, takes into account the fact that the sample is highly sheared before forming an expanding sheet. Our measurements have highlighted non-linear viscoelastic features that largely depend on the network structure. Note that in Refs.~\cite{Zhao1993,IgnesMullol1995}, analysis of the onset of fingering to fracturing transition above a critical injection rate in Hele-Shaw experiments in terms of a Deborah numbers based on the linear relaxation time seems to hold for some associative-polymer based samples, which form networks structurally similar to our microemulsion-based samples but clearly fails for shear-thinning materials made of semi-dilute polymer solution. These results are consistent with our findings.

Overall, our work evidences \textcolor{myc} {a peculiar instability process for a viscoelastic sheet} and highlights the crucial need to take into account the rheological nonlinear effects to properly characterize the behavior of expanding viscoelastic sheets.





\section*{Acknowledgements}
We acknowledge Laurent Bouteiller for provision of EHUT and Carole-Ann Charles for provision of some microemulsion-based samples.
This work was financially supported by People Programme (Marie Curie Actions) of the
European Union Seventh Framework Programme (FP7/2007-2013) Supolen under REA Grant
Agreement No. 607937, the labex NUMEV (ANR-10-LAB-20).



\balance


\bibliography{cracks} 
\bibliographystyle{rsc} 

\end{document}